\RequirePackage{fix-cm}
\documentclass{article}
\usepackage{graphicx}
\usepackage{setspace}
\usepackage{authblk}
\usepackage{booktabs}
\usepackage{color,soul}
\usepackage{subfigure}

 \usepackage{multirow}
 \usepackage[table,xcdraw]{xcolor}
 \usepackage[normalem]{ulem}
 \useunder{\uline}{\ul}{}

 \usepackage{glossaries}
\newacronym{AON}{AON}{Abbreviations Of Names}
 
\usepackage{geometry}
\begin{document}

\title{Sign Language Conversation Interpretation Using Wearable Sensors and Machine Learning}

%\author{Ziemowit Dworakowski\corref{cor1}}
%\ead{zdw@agh.edu.pl}
%\address{AGH University of Science and Technology, al. Mickiewicza 30, 30-059 Cracow, Poland }
\newcommand{\new}{\bullet}
\newcommand{\com}[1]{\textbf{\large\textcolor{red}{#1}}}	% For different color of the text (e.g. highlight changes)
\newcommand{\col}[1]{{\textcolor{blue}{#1}}}				% For clearly visible comments

\author{Basma Kalandar}
\author{Ziemowit Dworakowski}
\affil{\textit{AGH University of Science and Technology, Department of Robotics and Mechatronics}}
% Author Orchid ID: enter ID or remove command
%\newcommand{\orcidauthorA}{0000-0001-8383-4366} % Add \orcidA{} behind the author's name
%\newcommand{\orcidauthorB}{0000-0002-5289-9826} % Add \orcidB{} behind the author's name

\newcommand{\FSize}{0.3}	% So all figure sizes could be updated in one spot

\maketitle

\begin{abstract}

The count of people suffering from various levels of hearing loss reached 1.57 billion in 2019. This huge number tends to suffer on many personal and professional levels and strictly needs to be included with the rest of society healthily. This paper presents a proof of concept of an automatic sign language recognition system based on data obtained using a wearable device of 3 flex sensors. The system is designed to interpret a selected set of American Sign Language (ASL) dynamic words by collecting data in sequences of the performed signs and using machine learning methods. The built models achieved high-quality performances, such as Random Forest with 99\% accuracy, Support Vector Machine (SVM) with 99\%, and two K-Nearest Neighbor (KNN) models with 98\%. This indicates many possible paths toward the development of a full-scale system.

\end{abstract}


\setcounter{section}{0}

\section{Introduction}
\label{sec:Intro}

		\subsection{Addressing the problem}

\begin{figure}[]
\centering
%width=\FSize\textwidth
		\includegraphics[width=12cm]{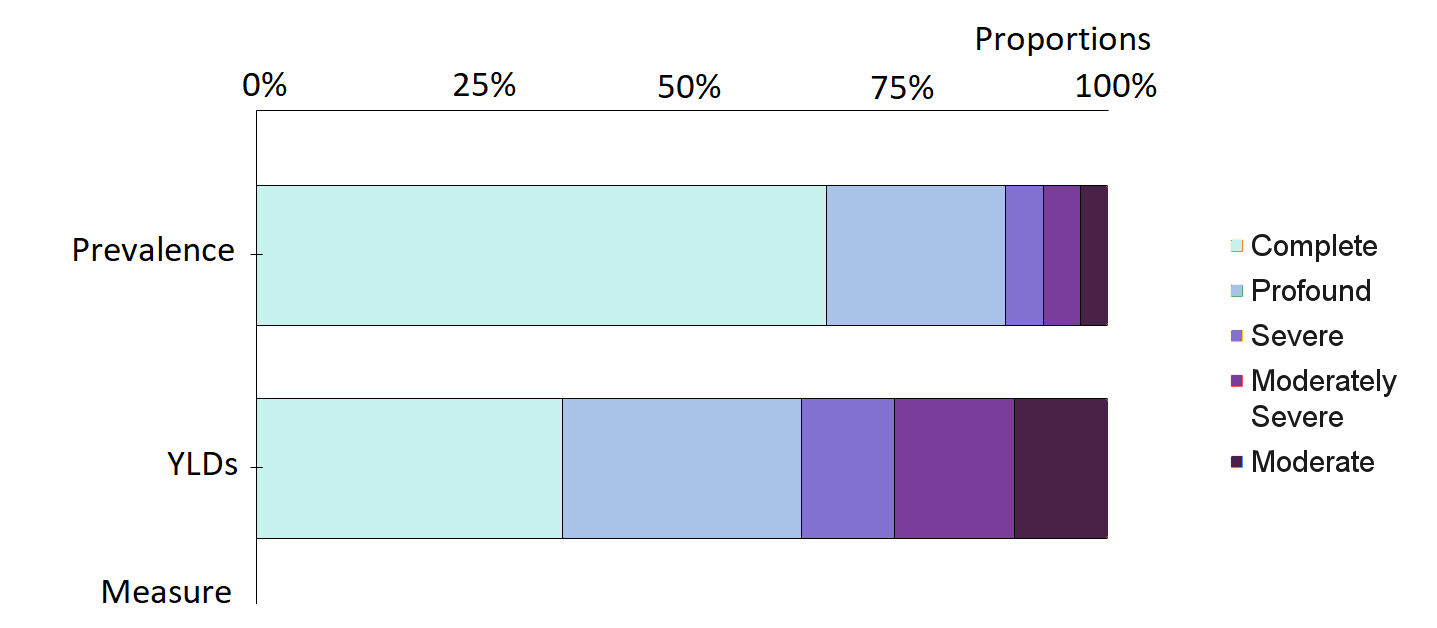}
		\caption[]{Proportion of individuals with moderate-to-complete
hearing loss by measure and severity}
		\label{fig:YLDfigure}
\end{figure}

People who suffer from different levels of hearing loss (impaired) and the disabled; all end up being partially, if not almost completely, excluded in their societies due to losing the possibility of communication through words. In 2019, 1.57 Billion people globally suffered from different levels of hearing loss \cite{Haile2021}. Figure \ref{fig:YLDfigure} shows the proportion of hearing loss levels among people between the prevalence of hearing loss and the number of Years Lived
with the Disability (YLD).

According to Fernandes et al. \cite{Fernandes2020} quoting the World Health Organization (WHO) in 2020, "a total of 466 million individuals globally have a hearing impairment, and this number is predicted to rise to 900 million individuals in another 30 years, by the year 2050. Furthermore, current estimates suggest an 83 percent gap in the available aid, i.e., only 17 percent of those who require aid can use it.". This means that 747 million individuals worldwide won't be provided with the necessary means to compensate for their disabilities. The fact that these people are being held back affects them, their families, their community, and society.\\

The communication quality for the impaired suffers mostly due to the lack of knowledge of sign language among common people. These negatively affect their employment chances, cognitive development, and emotional and mental well-being. \\ 

Communication for the impaired is based on sign language, which mainly depends on the hand (or both hands) shape, the hand position relative to the signer's body, and the motion done by the signer. Facial expression is an additional factor used to convey complementary meanings or intentions, such as emotions, agreeing and disagreeing, and indicating a question rather than a statement.

		\subsection{Related Work}
The need for a practical, reliable system to close the communication gap between the impaired and those around them is crucial on both individual and community levels. Automatic recognition of sign language is thus an important area of research. Modern approaches to address the problem focus mainly on the hands gestures previously mentioned, either the hand shape and position (known as static signs) or the hand shape, position, and movement (known as dynamic signs). As a result, they tend to omit the facial expression factor as it is not essential for the initial interpretation of the signs as much as giving them and the conversation more context, which is mostly intuitively understood by the addressed person. Those approaches are divided into two main categories:

\begin{enumerate}
  \item \textbf{Vision-based} Solutions based on this approach are camera oriented and focused on capturing the hands' shape or motion and isolating it from the rest of the frames. Image-based algorithms can directly process these frames before feeding them into a machine learning model, such as following the hand trajectory, which reached 99\% accuracy for 40 dynamic words in 10'000 images \cite{He2019}. Other examples of a similar method were reported to score  97.48\% \cite{Azar2020}, or 97.22\% accuracy \cite{Madani2013}. Histogram-based solutions tend to score lower results and require a more difficult setup, but can still achieve a good overall performance of 86\% \cite{Chen2016}.
Other popular vision-based approaches run the obtained frames through further processing. Leap Motion Controller (LMC), which can be directly used in real-time without an additional dedicated camera for optical hand tracking, obtains 96\% accuracy for static signs \cite{Naglot2016}, and 100\% for static numeral recognition from 0 to 9 \cite{Naglot}. Kinect SDK gets 3D skeleton information, which resulted in 95.68\% accuracy for dynamic words \cite{Xiao2020} and 94.74\% accuracy for dynamic words \cite{Naglot2016}. It is also possible to use sensor fusion with both LMC and Kinect SDK, like in the work by Kumar et al., to obtain 98.93\% accuracy for single-handed gestures and 95.92\% for double-handed gestures with 22 words used in the system \cite{Kumar2017a}. 
The main issue with vision-based solutions is the need for a predetermined environment, including camera range and angle setup. It is not practical in real-life situations with the dynamic events of the day because it takes out the natural flow of communication and requires a signer to be present in a particular area to communicate.

  \item \textbf{Wearable-device-based} Solutions in this category use a set of sensors attached to the signer's body to capture the motion and interpret the signs \cite{Zazoum2022} .Examples vary in technology and complexity including a solution using a smartwatch and EMG sensors band which interpreted 200 scentences containing different mixes of 100 words with accuracy reaching 99.3\% \cite{Zhang2022_zd} obtaining the highest accuracy found so far in the dynamic sentence interpretation using a wearable device, Inertial Measurement Units (IMU) which are used to calculate the trajectory and the angular position of the hand usually combined with other sensors. Such a method can obtain 93\% accuracy for 40 static words \cite{Ahmed2021}, 96\% for the alphabets (static), and 99\% for static numerical values with a system containing 17 sensors, or 96\% accuracy for 7 dynamic gestures and obtaining 95\% average accuracy using the system of 5 flex sensors, one 3-axes accelerometer and 3 FSR sensorsto interpret words \cite{Chu2021}. Electromyography (EMG) based sensors read the body muscle changes reaching 96.5\% accuracy for 120 Chinese subwords and 86.7\% for 200 Chinese sentences using 8 EMG sensors and 2 accelerometers \cite{Li2012} and in another article \cite{Zheng2022_zd} that used MYO bracelet containing 8 EMG sensors, a 3-axes accelerometer and a 3-axes gyroscope achieved 90.05\% accuracy in recognizing 200 words. Photoplethysmography (PPG) manipulated to reads the change in the blood volume through the signer's wrist as they sign reaching 98\% accuracy for static gestures \cite{Zhao2021}. Force sensors are reported to produce 91.11\% accuracy for static words \cite{Al-Hammouri2021}.\\

\end{enumerate}

%\singlespacing
%\scriptsize

\begin{table}[]
\centering\scriptsize
\caption{Existing Solutions}
\begin{tabular}{p{1cm}|p{2cm}|p{2cm}|p{2cm}|p{2cm}|p{2cm}} \toprule
\textbf{Citation}	
&\textbf{Dataset Size}	
&\textbf{Type of Data}	
&\textbf{Method}	
&\textbf{System size}	
&\textbf{Maximum Accuracy}		\\ \toprule

\cite{Fernandes2020}	&31'200	&26 Static signs 	&Vision based	& - &99.98\%	\\ 
\cite{He2019}		&10'000	&40 words		&Vision based	& - &99.0\%	\\
\cite{Azar2020}		&1200		&20 words		&Vision based	& - &98\%	\\
\cite{Madani2013}		&600		&20 words		&Vision based	& - &95.56\%	\\
\cite{Chen2016}		&26200	&72 words (22 one handed gestures)	&Vision based (kinect) & - 	&89.6\% (98.2\% one handed gestures)	\\

\cite{Naglot2016}, \cite{Naglot}		&520		&ASL alphabet (26)	&Vision based (leap motion) & - 	&96.15\%	\\

\cite{Xiao2020}		&12'500	&50 words		&Vision based	& - &95.86\%	\\
\cite{Kumar2017a}		&840 (single hand)	&		&Vision based	& - &98.93\%	\\
\cite{Kumar2017a}		&660 (both hands)	&50 words(combined)	&Vision based	& - &95.91\%	\\
\cite{Zhang2022_zd}		&20'000	&Dynamic 250 sentences with 100 words	&Wearable device	&Smart watch and sensors armband	&99.3\% \\
\cite{Zheng2022_zd}		&4000		&200 words			&wearable device	&10 sensors (MYO)	&90.05\% \\
\cite{Ahmed2021}		&1875		&75 static words	&Wearable device	&17 sensors	&93.4\%	\\

\cite{Chu2021}		&2100		&7 words		&Wearable device	&9 sensors	&96.9\%	\\
\cite{Li2012}		&13'920	&120			&Wearable device 	&10 sensors	&96.9\% \\
\cite{Zhao2021}		&7000		&static letters \& numbers&Wearable device	&5 sensors	&98\% \\
\cite{Al-Hammouri2021}	& -		&static letters \& numbers&Wearable device	&6 sensors	&91.11\%

										\\ \bottomrule
\end{tabular}
\label{tab:table2}
\end{table}	

Table \ref{tab:table2} summarizes examples of research within both categories. In most cases, the problem tackled within the reported papers was to classify a selected subset of the whole sign language. Such a subset can cover several expressions, numbers, or letters. Still, the reported results are usually much lower than 99\%. Unfortunately, the scalability of the reported solutions is challenging, as the problem complexity tends to rise exponentially with the increase of the number of classes to recognize. It is, therefore, crucial to design new solutions focusing on their scalability, with the end-goal being to reach the recognition of all of the expressions of the sign language in one decision system. To this end, the focus should now be put on finding solutions that can reach high sign recognition rates using as little hardware and sensor complexity as possible; as a result, ensuring space for further scalability of the system by adding more sources of information to recognize new gestures with a large safety margin.

The work presented in this paper falls within the wearable-device-based category. It consists of a scalable, low-cost, low-complexity system that incorporates just three flex sensors while already resulting in a very high overall performance. Data recorded by the system is used to compare five different decision algorithms, namely: Random Forest (RF), the k-Nearest Neighbor classifier (KNN), a Decision Tree (DT), Logistic Regression (LR), and Support Vector Machine (SVM). The comparison is made based on a 23-dynamic-gesture database.
Solutions in this category use a set of sensors attached to the signer's body to capture the motion and interpret the dynamic expressions.

		\subsection{Flow of Sequence}

In the section 2 the methodology's details  for the work approach are provided, including the signs setup, data collection method, the hardware used, software concepts and classifier types. Section 3 contains the experimental setup, conditions and assumptions as well as the training models and the parameters chosen to measure the models' performances with the testing data. Section 4 provides discussion, conclusions and potential directions for further development of the method.

%%%%%%%%%%%%%%%%%%%%%%%%%%%%%%%%%%%%%%%%%%

\section{Materials and Methods}
\label{sec:Method}

		\subsection{Method assumptions}	
The system chosen for the sign language interpreter is oriented around fulfilling two core concepts:

\begin{enumerate}
  \item \textbf{Practicality, user accessibility, and convenience}: The system must simplify the signer's communication by being conveniently accessible at all times without depending on external setups and circumstances.
  \item \textbf{System simplicity with the highest performance}: Since this is a proof of concept and doesn't contain the entire sign vocabulary of ASL, the aim is to make the complexity the least in both the hardware and the software of the system. 
\end{enumerate}

These concepts resulted in the following list of constraints for the final method:
	
\begin{itemize}
  \item The system is designed around interpretation of ASL only.
  \item For simplicity, all ASL signs can be captured through 3 sensors only.
  \item The system interprets words and expressions instead of the more common approach of letter interpretation.
  \item All signed expressions take the same amount of time to be done.
  \item A set of 23 words and expressions is chosen as a subset of the entire ASL - to demonstrate system's capabilities.
  \item Data collection is done in the form of arrays of instances, not a single entry nor a single value extracted for a set of values.
  \item The device is fitted to the user prior to data capture. Separate training for a new user is not required
\end{itemize}

		\subsection{The Language Setup}

The system setup is performed based on a dataset containing 23 different dynamic signs. The chosen expressions are presented in Table \ref{tab:table1} to show the notation used to represent each sign and its corresponding meaning.

\begin{table}[]
\centering\scriptsize
\caption{Vocabulary used}
\begin{tabular}{p{3cm}|p{3cm}} \toprule
\textbf{Notation}							&\textbf{Word}	\\ \toprule
hello								& Hello			\\ 
welcome								&Welcome		\\ 
hru									&How are you?	\\ 
canIHelpU							&Can I help you?\\
whatsup								&What's up 		\\
busy								&Busy			\\
nothing								&Nothing		\\
yes									&Yes			\\
no									&No				\\
deaf								&Deaf			\\
hardHearing							&Hard of hearing\\
learn								&Learn			\\
ASL									&ASL			\\
want								&Want			\\
sorry								&Sorry			\\
please								&Please			\\
CULater								&See you later	\\
ok									&OK				\\
notALot								&Not a lot		\\
signLanguage						&Sign language	\\
have								&Have			\\ \bottomrule
\end{tabular}
\label{tab:table1}
\end{table}

		\subsection{Hardware Setup}
Since the focus of the paper is to present research work on new ways of data processing and interpretation, the hardware setup is not production-ready and is prepared as a stationary laboratory setup based on Arduino architecture. The system consists of three Arduino Flex sensors (2 of 112*6.3mm and 1 of 73*6.3mm) and three constant resistors, one for each sensor, used as voltage dividers connected to the Arduino microcontroller using a breadboard and jumper wires.
The elbow, thumb, and middle finger joints are chosen for the proof of concept. The elbow is included in the system because, in the performing of dynamic signs, the hand's shape can be similar in multiple words, but the extension or motion of the elbow differentiates between those signs. The vocabulary selected for this work focuses on words that would include at least one of those joints' distinct motions.\\

\begin{figure}[]
\centering
%width=\FSize\textwidth
		\includegraphics[width=12cm]{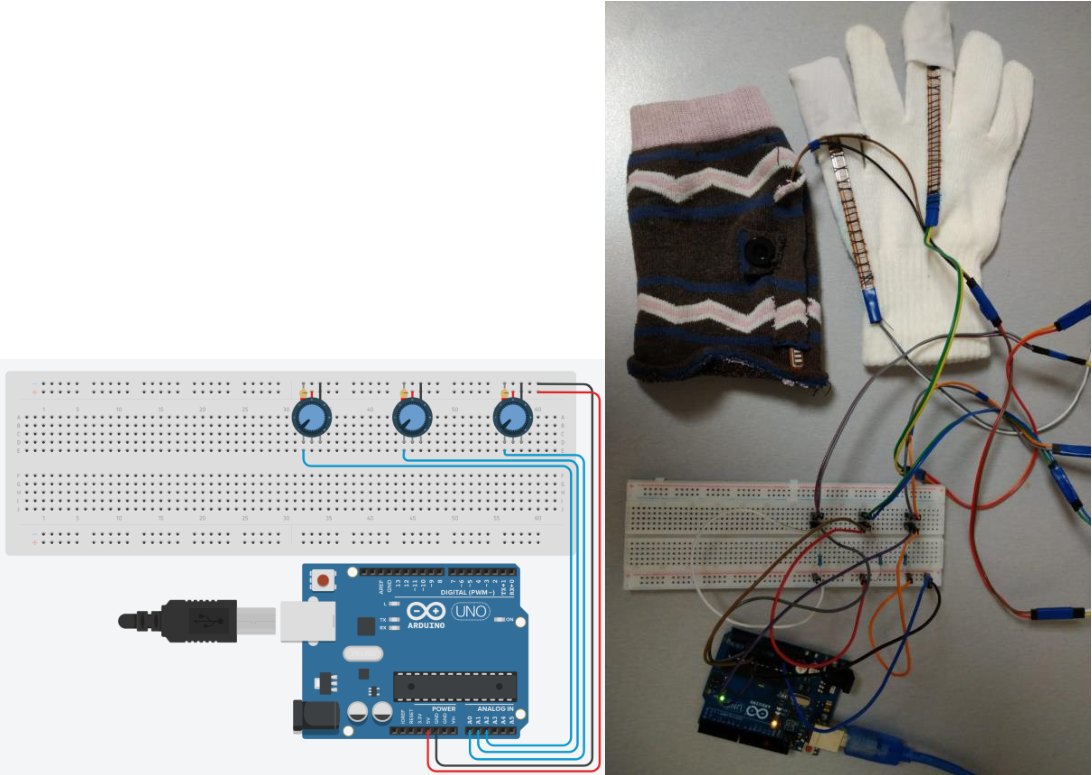}
		\caption[]{The hardware setup with (a) schematic of the system, where the potentiometers represent the flex sensors, and (b) in practical implementation, showing how the sensors are sewn into wearables.}
		\label{fig:circuitFig}
\end{figure}

Figure \ref{fig:circuitFig} shows the hardware's schematic and practical implementation.

		\subsection{Software setup}	
\begin{figure}[]
\centering
%width=\FSize\textwidth
		\includegraphics[width=12cm]{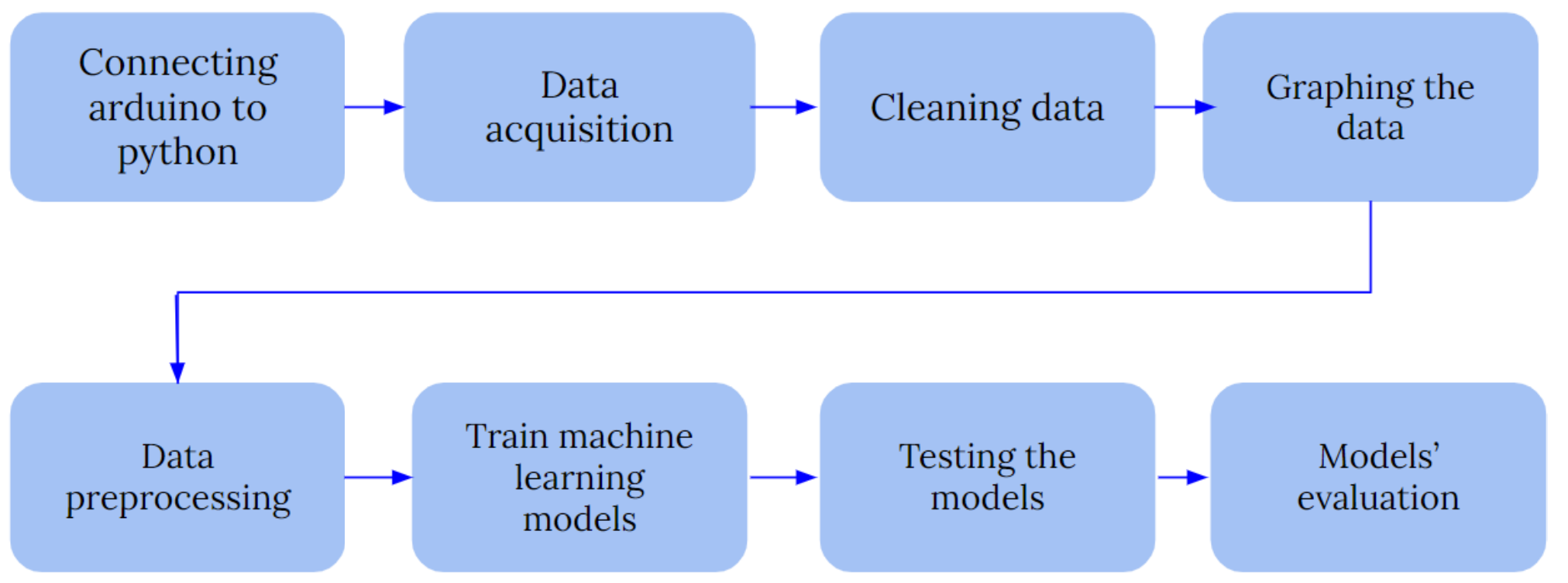}
		\caption[]{The block diagram for the sign language recognition system.}
		\label{fig:blockDia}
\end{figure}

The block diagram in figure \ref{fig:blockDia} demonstrates the sequence of the software system of the interpreter from the beginning of the connection to the models' results.

\begin{figure}[]
\centering

		\includegraphics[width=12cm]{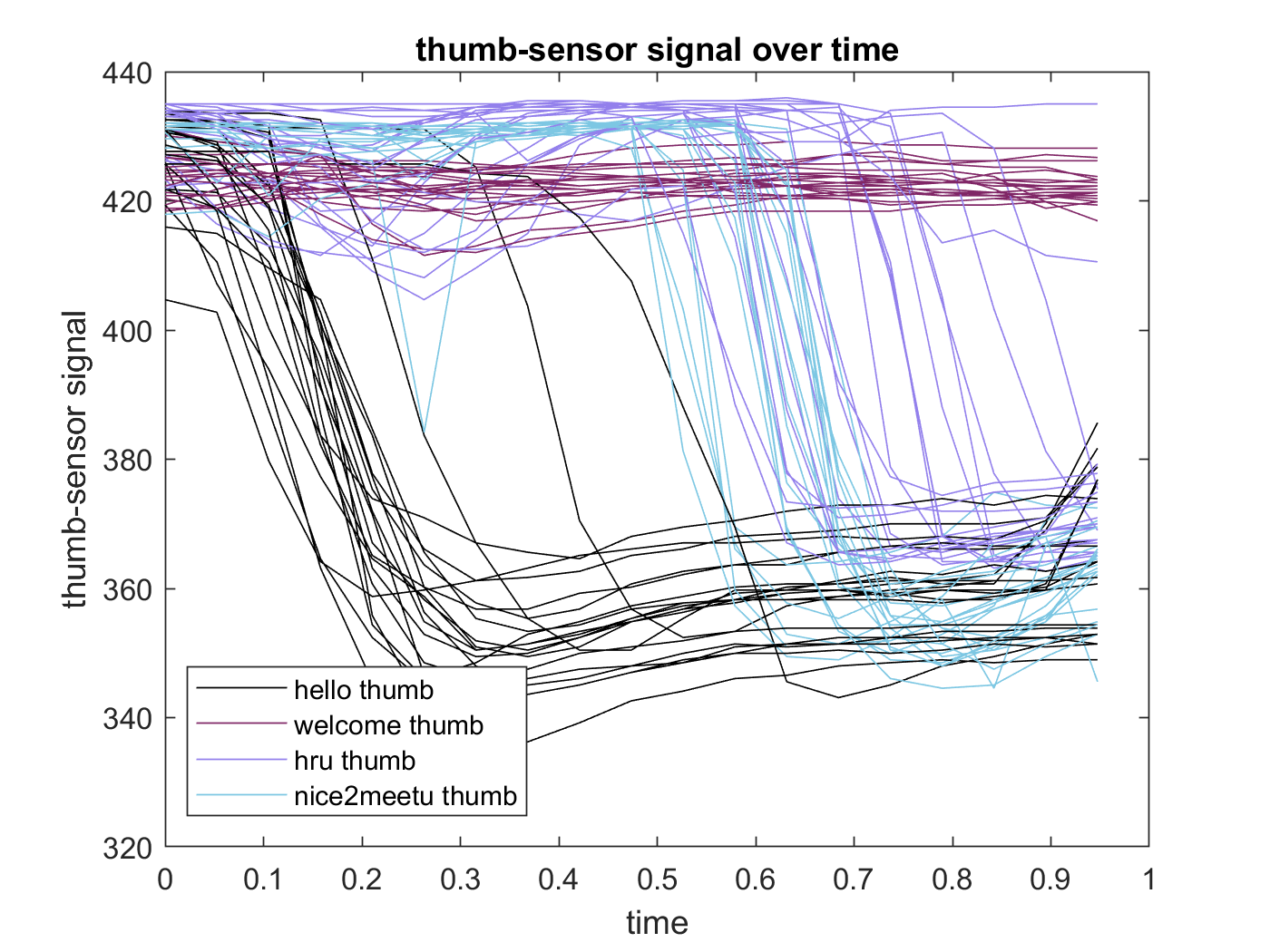}
		\caption[]{The thumb sensor data for 4 words accross time.}
		\label{fig:thumb}
\end{figure}

\begin{figure}[]
\centering
		{\includegraphics[width=0.4\textwidth]{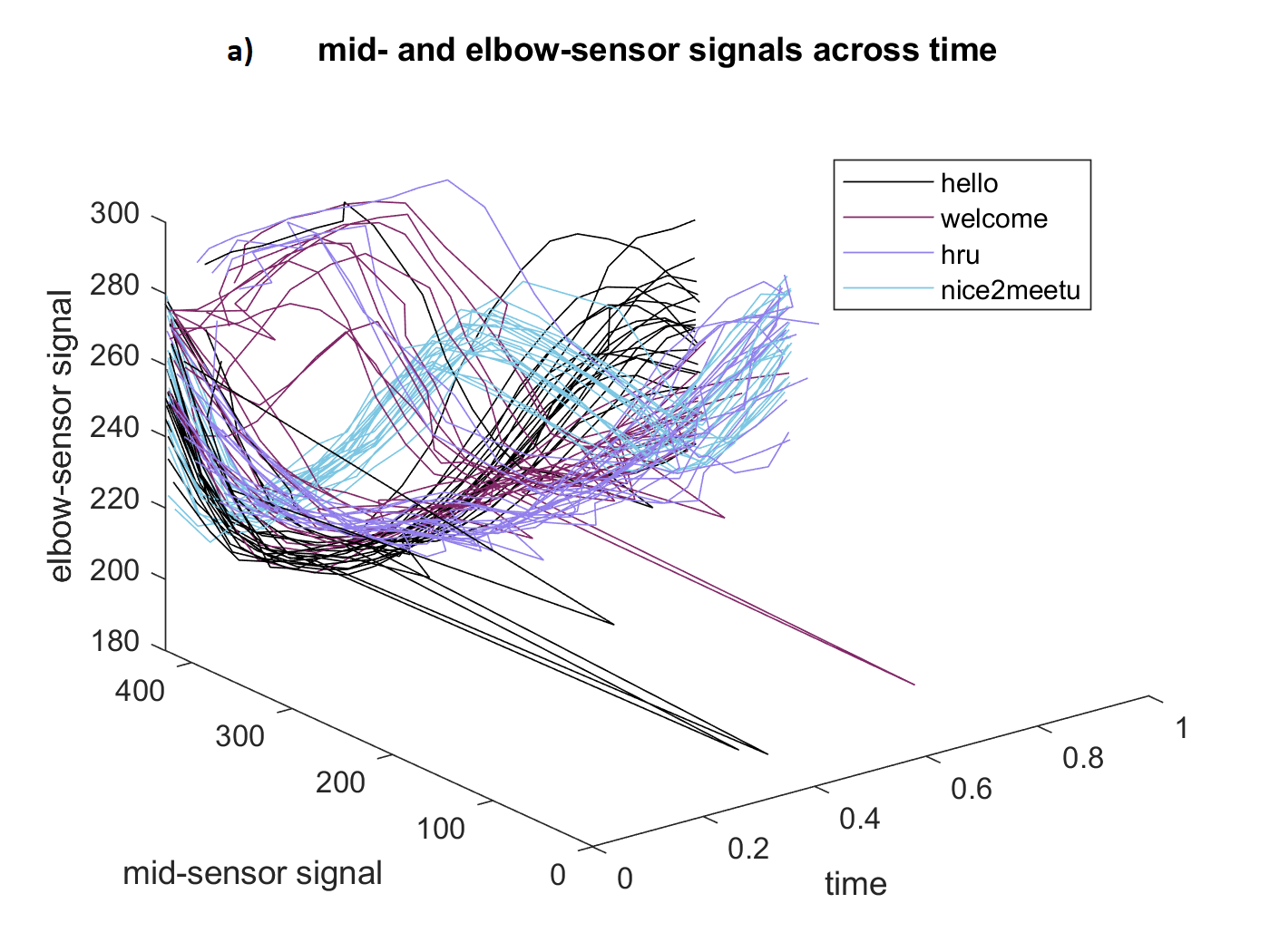}}
		{\includegraphics[width=0.4\textwidth]{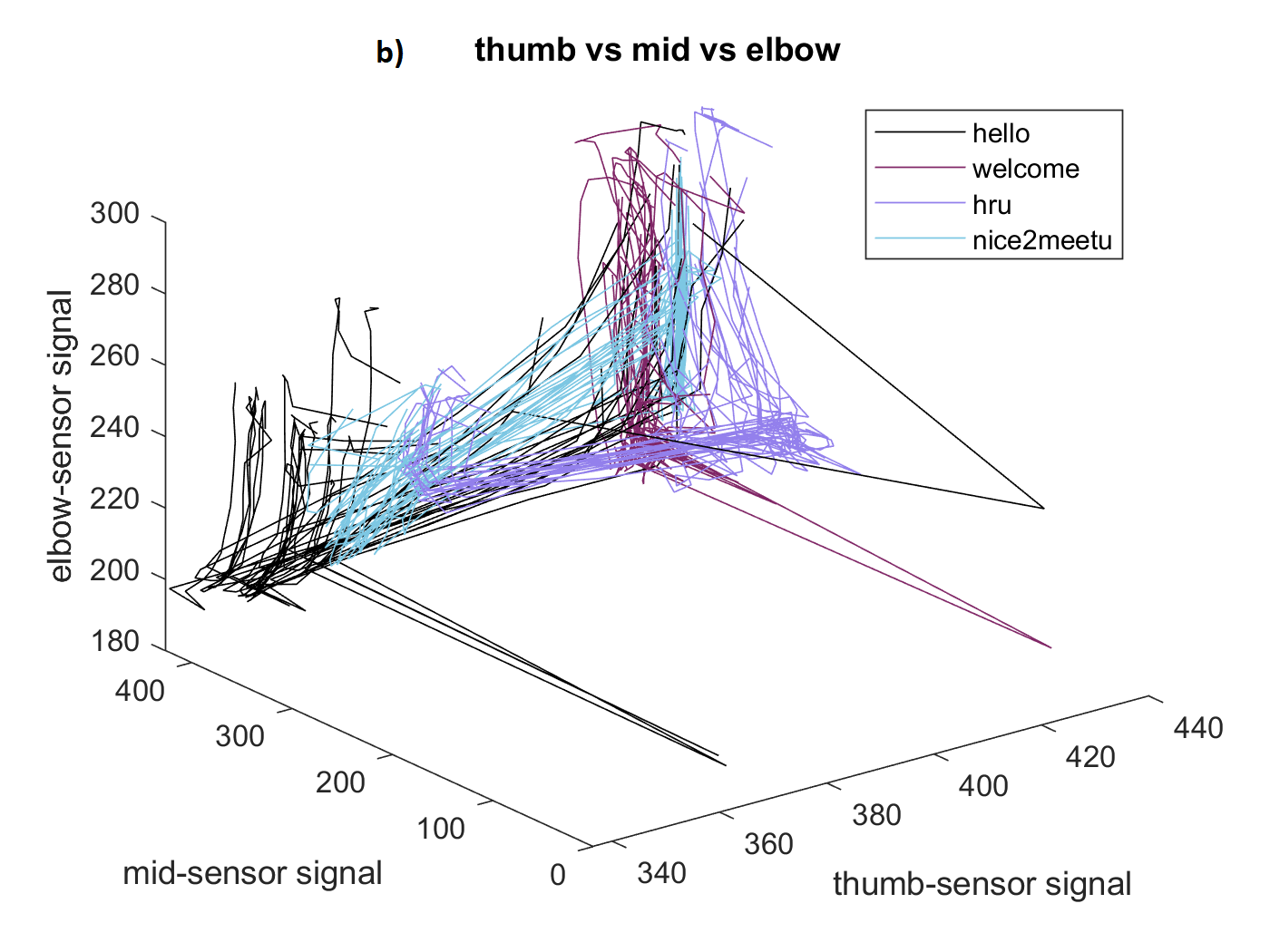}}\\
		{\includegraphics[width=0.37\textwidth]{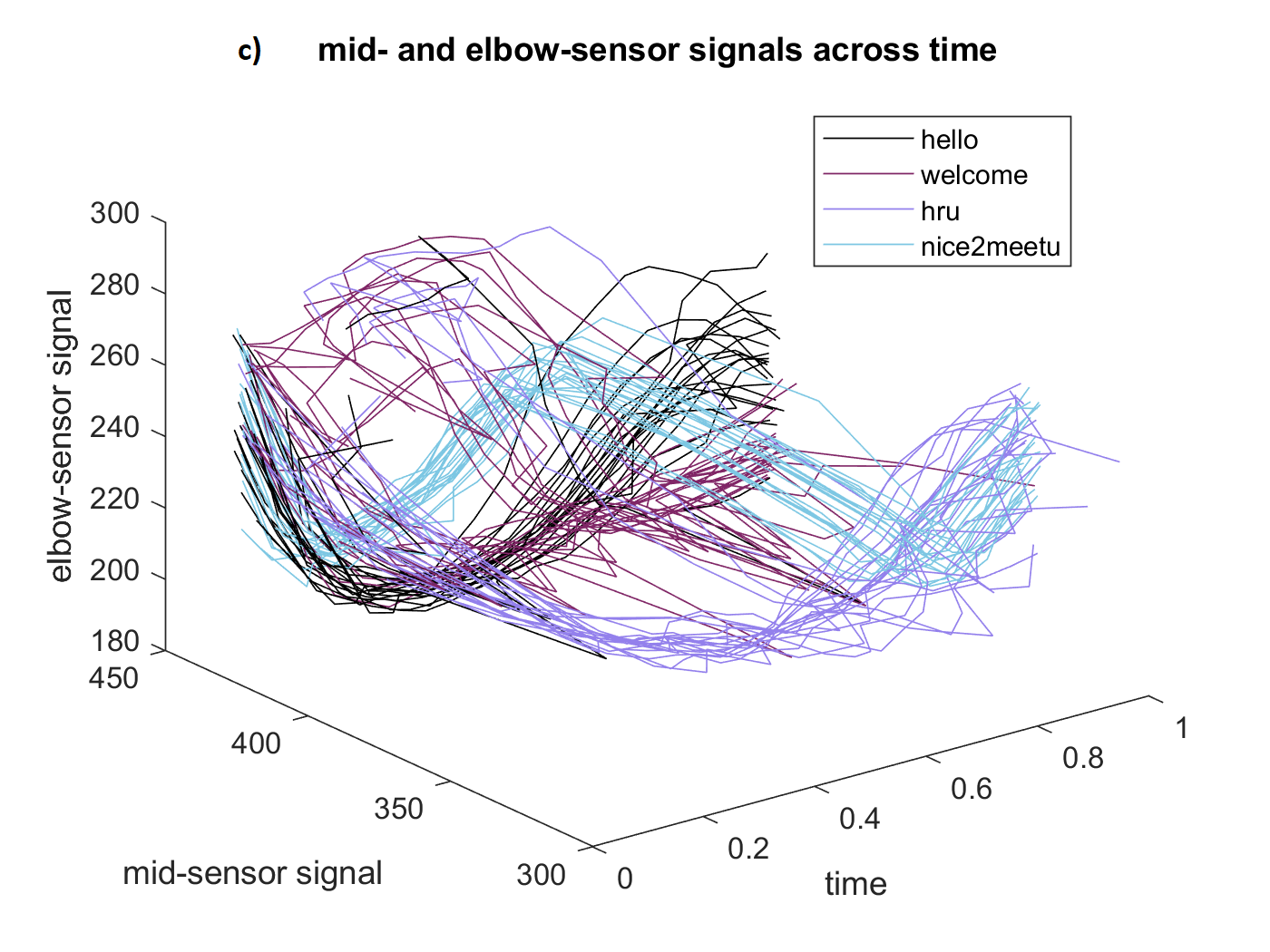}}
		{\includegraphics[width=0.4\textwidth]{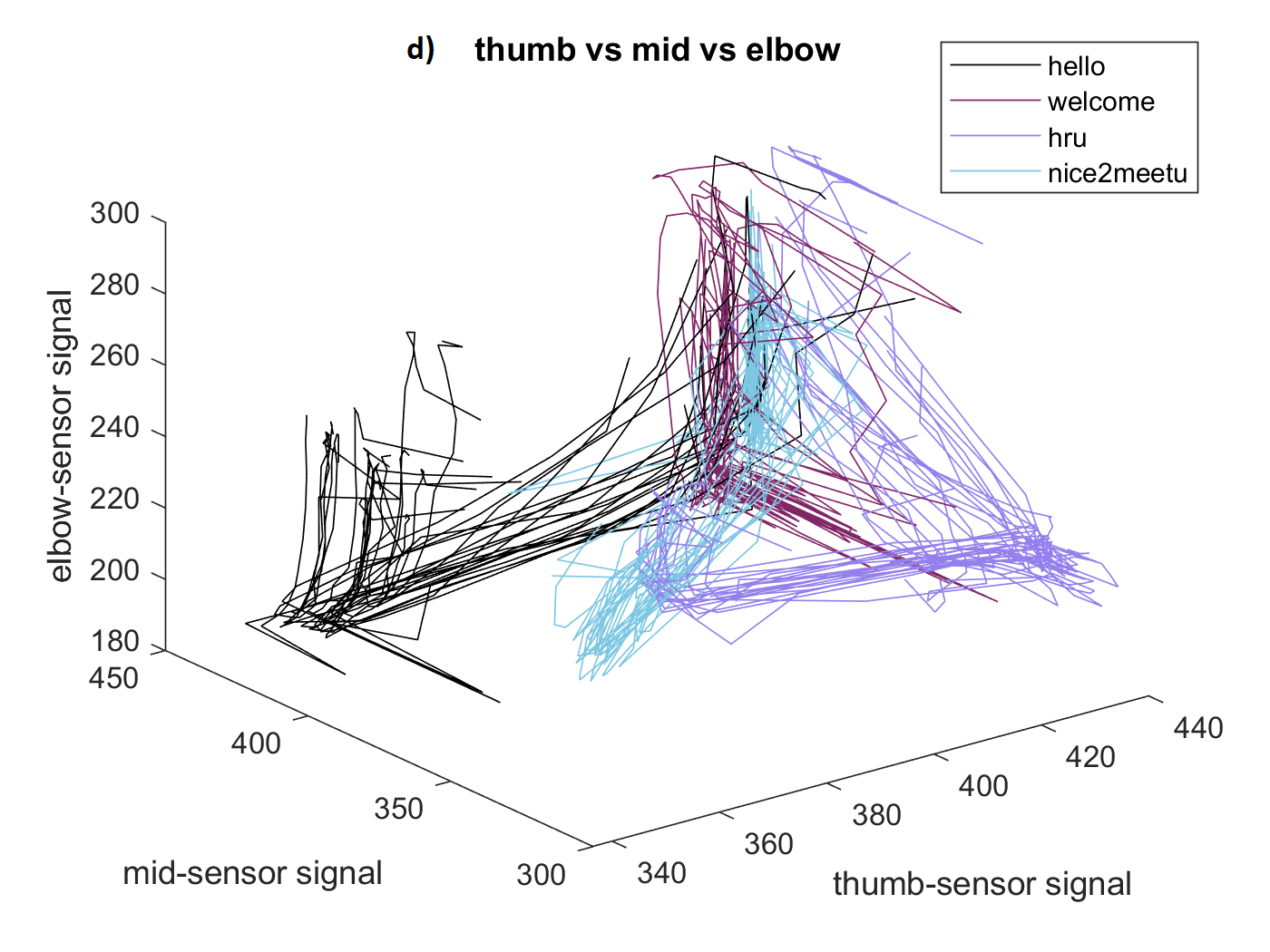}}
		\caption[]{Three-dimensional plots of data for (a) middle finger and elbow sensors in time-domain, raw data (b) trajectories build from three sensors, raw data, (c)  middle finger and elbow sensors in time-domain, cleaned data and (d) trajectories build from three sensors, cleaned data}
		\label{fig:graphsFig}
\end{figure}

The data acquisition method doesn't focus on snapshots taken during a specific sign moment. Instead, it captures the entire gesture into an array of inputs. A decision model would then learn the whole time-domain pattern combination of the three sensors for each word and classify it accordingly. Similar to vision-based approaches that are trajectory-based \cite{Azar2020}, the wearable device allows arrays of similar inputs that build unique, distinct, independent pattern combinations for each expression. Figure \ref{fig:thumb} shows the recording of thumb-related flex sensor over time for different gestures. Although some similarities are visible, like in the case of \textit{hru} (how are you) and \textit{nice2meetu} (nice to meet you) gestures, it is, however, not a problem as this is only true for two-dimensional representation. Representations that include all three sensors in the time-domain or even just three-dimensional trajectories without time factor present clear differences (Compare with Figure \ref{fig:graphsFig} (a) and \ref{fig:graphsFig} (d) \\

The next step after data acquisition is data cleaning. Its purpose is to delete momentary discontinuities and spikes. The procedure requires threshold-based check of registered signals. The spikes that breach the predetermined thresholds are automatically replaced with mean values of their neighbors. Signals that include longer periods of discontinuity are marked for manual investigation and deletion.\\

Figure \ref{fig:graphsFig} shows 4 dynamic signs. Graphs \ref{fig:graphsFig} (a) and \ref{fig:graphsFig} (c) demonstrate data acquired by sensors on the elbow and middle finger across time before and after cleaning. In Figure \ref{fig:graphsFig} (b) and \ref{fig:graphsFig} (d) the trajectories built from signals obtained from three sensors is presented. Again, it is clearly seen that the cleaning procedure eliminates most of the spikes and that the gestures significantly differ from each other.

Next, the data are combined to form the dataset fed for training the machine learning models for sign classification. Multiple models are designed to be tested and compared against each other, namely: LR, SVM, RF, DT and two KNN models. 

\section{Experimental evaluation}
\label{sec:expSetup}

		\subsection{Experimental setup}
The device is worn by the user and remains on standby until the user begins to sign, then it starts the system starts recording the data. Each sign is split into 19 instances using three sensors, which add up to 57 scalar values per sign. Because of the small scale of the system, the models were created to take the whole sign data as input. 

A dataset of 1044 words and expressions (after data cleaning) is created using the device. The dataset is shuffled, then approximately 80\% of this dataset (that is: 835 words and expressions)- is used for training the classifiers. The remaining samples are used for the performance test of the final classifiers.\\

The machine learning models are done in python using \textit{sklearn} package and jupyter notebook in anaconda3. The classifiers' hyperparameters were either left at their default values or modified slightly based on authors' expertise and trial-and-error approach. The final set of hyperparameter values is as follows:

\begin{enumerate}
  \item{ \textbf{RF}: RandomForestClassifier(n\_estimators=100, max\_depth=10, random\_state=0)}
  %\item \textbf{SVM}: make_pipeline(StandardScaler(), SVC(gamma='auto'))
  \item \textbf{KNN Model 1}: KNeighborsClassifier(n\_neighbors=5)
  \item \textbf{KNN Model 2}: KNeighborsClassifier(n\_neighbors=3)
  %\item \textbf{Logistic Regression}: LogisticRegression(warm\_start = True, solver = 'sag',  random_state = 42, multi_class = 'multinomial', max\_iter=1000)
  \item  \textbf{DT}: DecisionTreeClassifier(random\_state=0) 
\end{enumerate}

		\subsection{Evaluation criteria}
The classifiers evaluation is statistic-based.

\begin{itemize}
  \item\textbf{Accuracy}: the model's ability to correctly classify data.
  \item\textbf{Precision}:  the ratio of the correctly classified samples in a class among all the samples classified in that class.
  \item\textbf{Recall}: the ratio of correctly classified samples in a certain class with respect to all data points belonging to that class.
  \item\textbf{F1 Score}: Correct positive predictions percentage.
  \item\textbf{Confusion Matrix}: to view the classification performance details for all the data points
\end{itemize}

The weighted average of these parameters are the focus of the evaluation because that way if there are any imbalances in the tested data, they will be taken into account.

		\subsection{Experimental results}

After testing all the trained classifiers, a comparison between them is drawn in Table \ref{tab:table3} showing the weighted averages of the accuracy, precision, recall and f1 score.

\begin{table}[]
\centering\scriptsize
\caption{Vocabulary used}
\begin{tabular}{p{1cm}|p{1cm}|p{1cm}|p{1cm}|p{1cm}|p{1cm}|p{1cm}} \toprule
\textbf{Criteria}	
&\textbf{RF}	
&\textbf{SVM}	
&\textbf{KNN (k=5)}	
&\textbf{KNN (k=3)}	
&\textbf{LR}	
&\textbf{DT}		\\ \toprule

Accuracy	&99\%	&99\%	&98\%	&98\%	&93\%	&90\%	\\ 
Precision	&99\%	&99\%	&98\%	&98\%	&94\%	&90\%	\\ 
Recall		&99\%	&99\%	&98\%	&98\%	&93\%	&90\%	\\ 
F1 score	&99\%	&99\%	&98\%	&98\%	&94\%	&90\%	\\ 
										\\ \bottomrule
\end{tabular}
\label{tab:table3}
\end{table}	

\begin{figure} []
\centering
		{\includegraphics[width=0.4\textwidth]{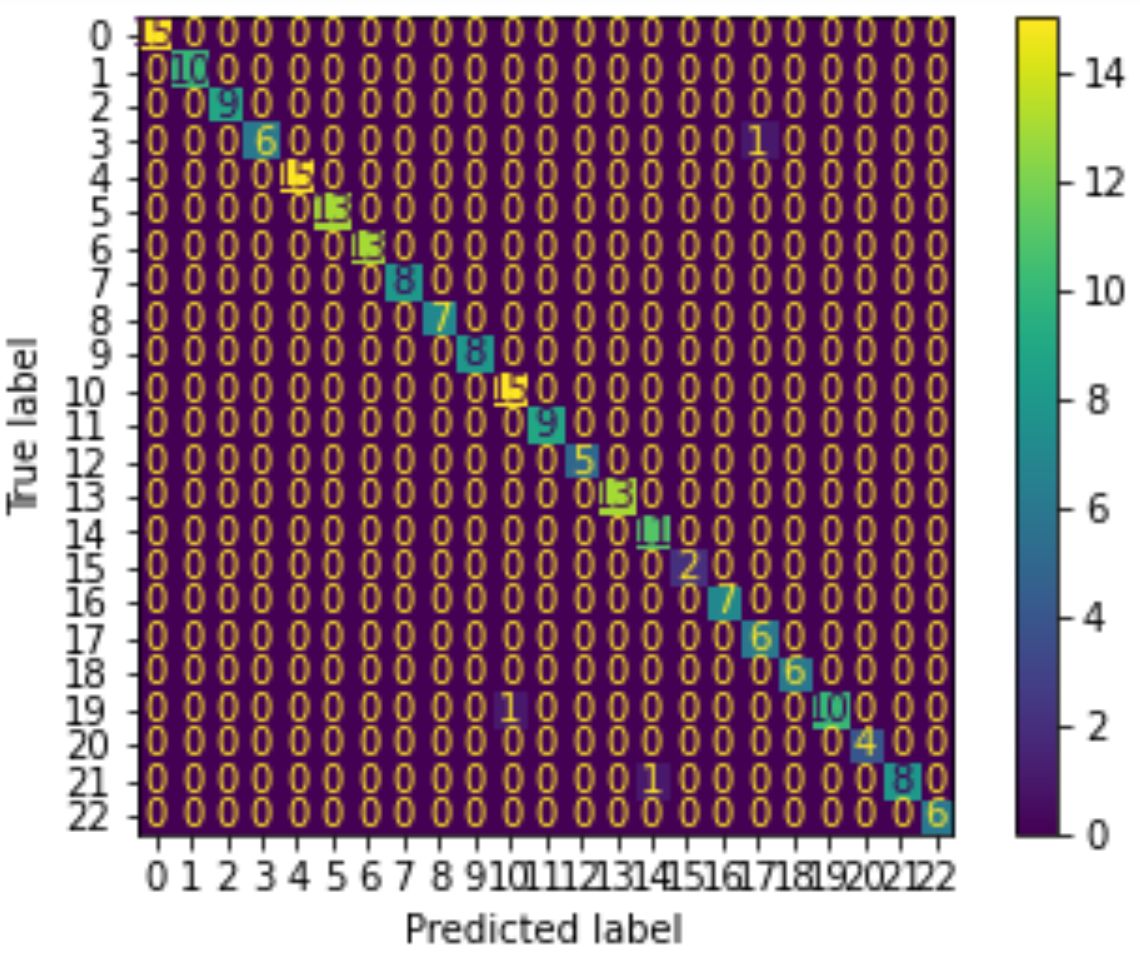}}
		{\includegraphics[width=0.4\textwidth]{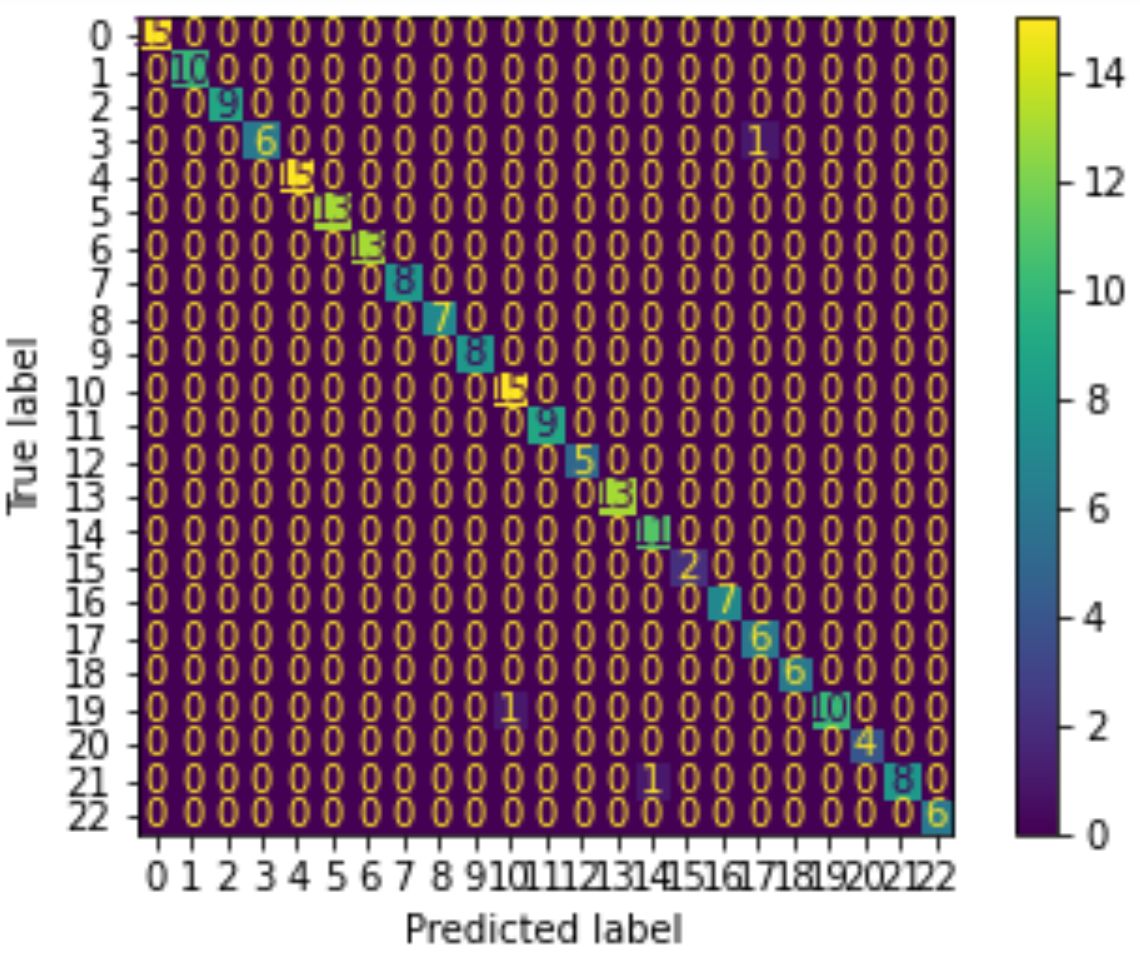}}\\
		{\includegraphics[width=0.4\textwidth]{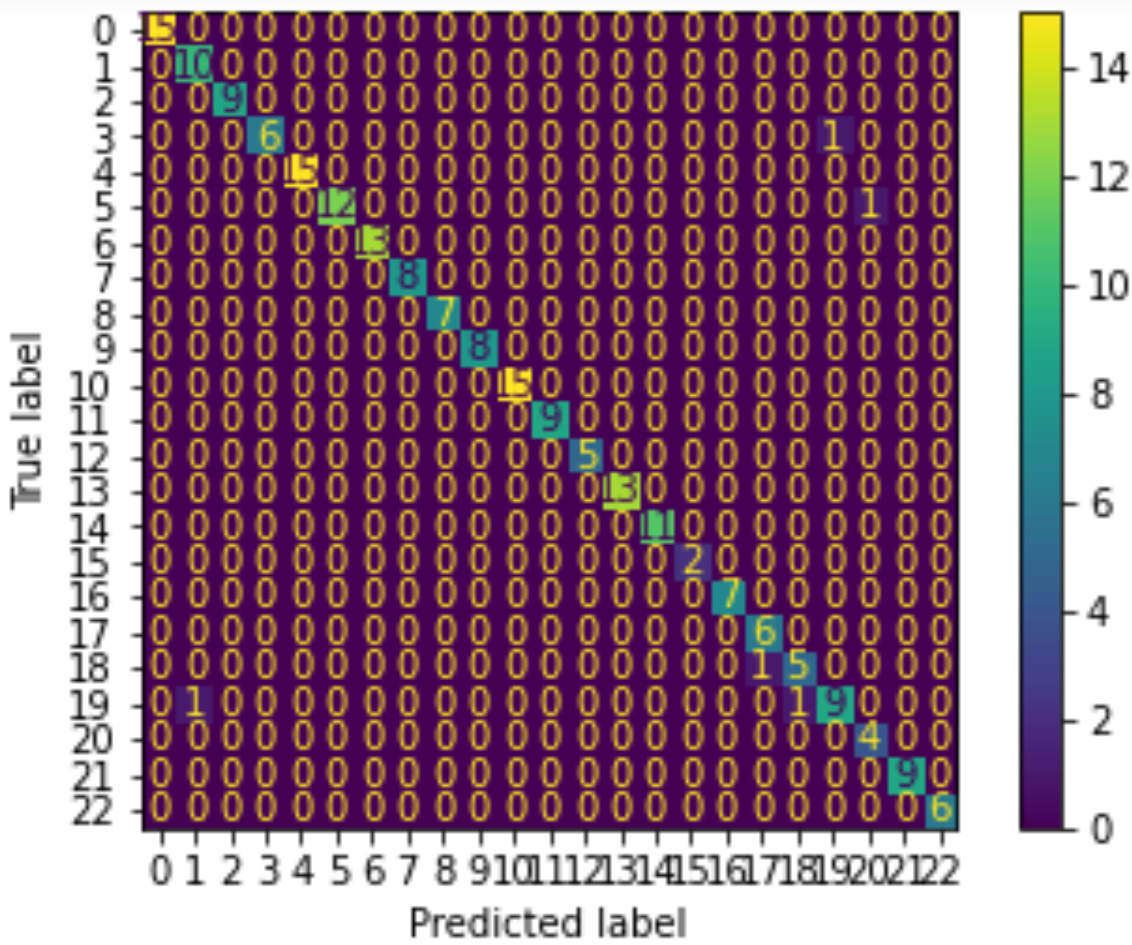}}
		{\includegraphics[width=0.4\textwidth]{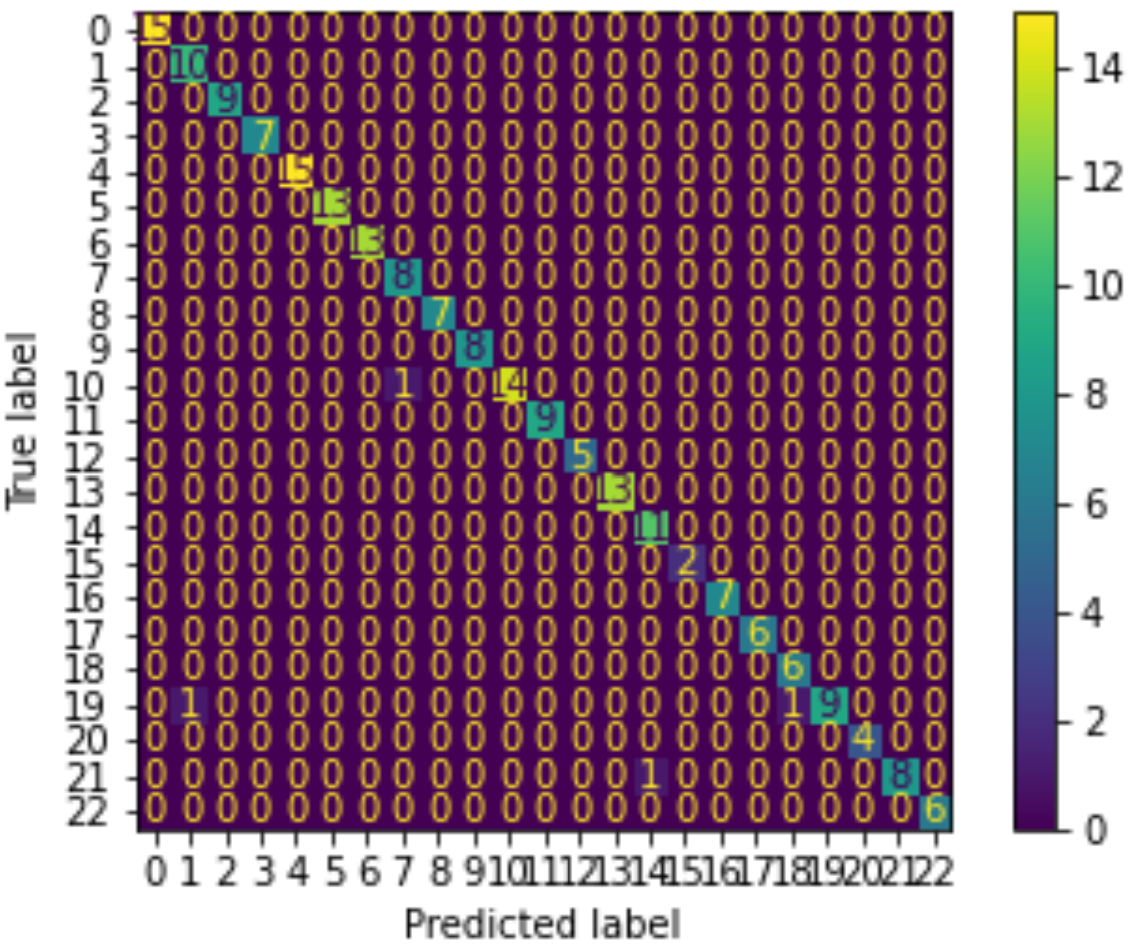}}\\
		{\includegraphics[width=0.4\textwidth]{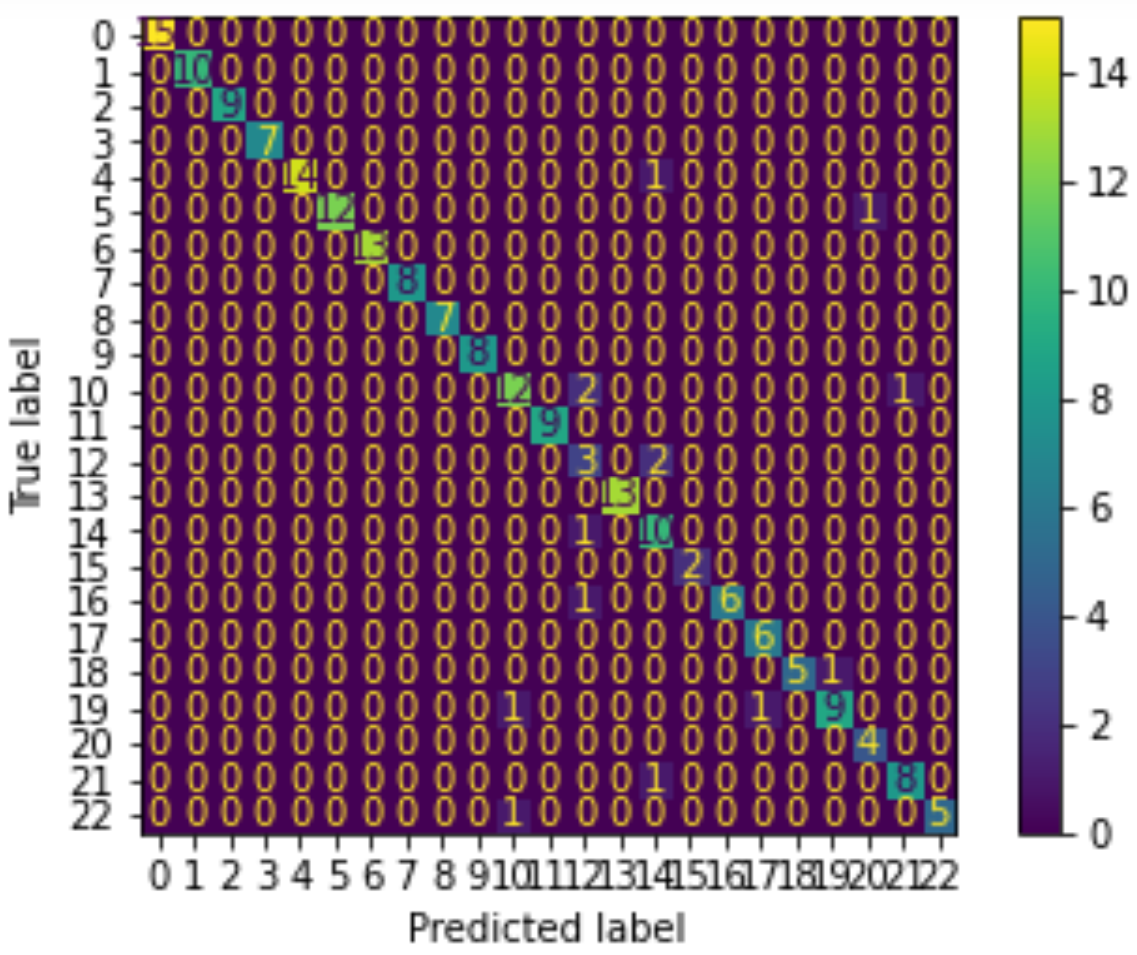}}
		{\includegraphics[width=0.4\textwidth]{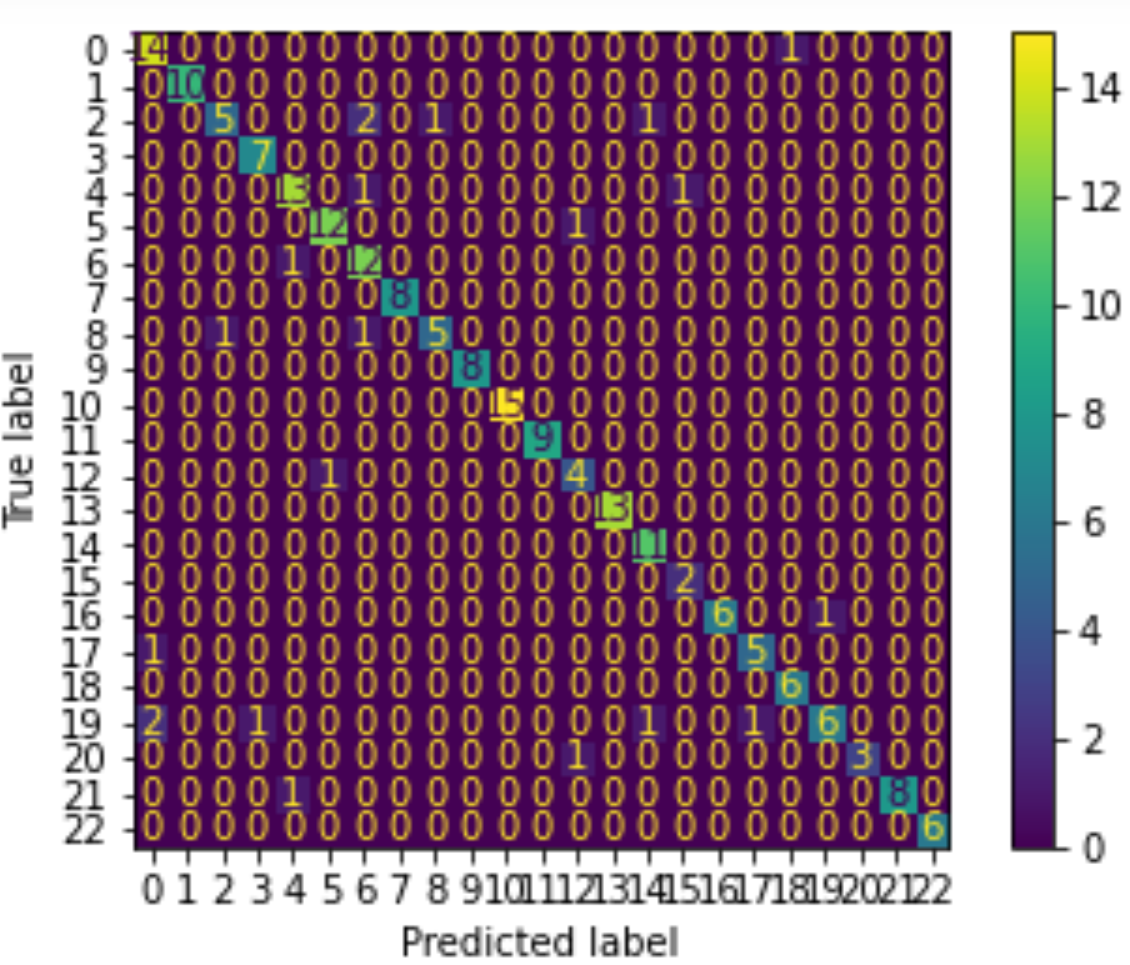}}\\
		\caption[]{Confusion matrices of (a) RF, (b) SVM, (c) KNN -k=5-, (d) KNN -k=3-, (e) LR, (f) DT}
		\label{fig:cm}
\end{figure}

Figure \ref{fig:cm} shows the confusion matrices for all the classifiers. They don't show patterns of misclassifying specific signs nor random overall failure in the system.

%%%%%%%%%%%%%%%%%%%%%%%%%%%%%%%%%%%%%%%%%%
\section{Discussion}
\label{sec:disc}

		\subsection{Results indications}
All the obtained accuracies are in the range of 90\%  to 99\%. This reflects both a successful data acquisition method (from hardware setup to the method of collecting data) and well-constructed and fitted models to the provided dataset of 23 words despite using only three sensors for data acquisition.
The confusion matrices showed that the misclassifications are distributed evenly in the whole matrix. Thus, they show there are no particular problematic words. This shows that even with only three sensors, the data collected are distinct enough from each other.\\

The method of collecting arrays of time-domain data instead of a snapshots of sensor readings at a particular moment, following with a full-data analysis instead of feature calculation was proven to be efficient. The machine learning models' hyperparameters were mostly set at default values because the data acquisition alone allowed for obtaining easily separable data. This kind of simplicity contributed to minimizing errors for the final system. Another way to reduce potential errors in the current system, with the same training models, would be to increase the system's resolution within the same time limit, increasing the size of a data batch passed for evaluation. \\

		\subsection{Scaling up expectations}

The expectations for the models' performances with scaling up the method are not all positive:

\begin{itemize}
  \item\textbf{RF}: Shows good promise as it misclassified only two signs out of 209 in the testing process.It also uses many hyperparameters that allow for dealing with outliers, while the ensemble approach allows for maintaining high reliability. While they may prove challenging in the configuration in the long run, they certainly have high expected performance for large-scale systems. 
  \item\textbf{SVM}: The classifier's kernel can use higher degree polynomials which can help in the non-smooth classification, and the classifier's type holds out in higher dimension problems, so the increase of the number of sensors shouldn't create an issue for the classifier which makes it more suitable for the task.
  \item\textbf{KNN}:  expected to perform worse with the dimensionality increase because it will include more parameters for each dimension, and distance-based classifiers are known to scale poorly for multidimensional spaces. The point behind using KNN in the proof of concept is to prove that both KNNs, despite their simplicity, obtained 98\% of performance with dynamic motion recognition of ASL.
  \item\textbf{LR}: While its performance is not exactly bad with the provided parameters, it is expected to perform worse when the system's complexity increases. The classifier was already the slowest in the training process. In addition, the system never converged even after increasing the maximum number of iterations to 1000. It only slowed the system and increased the performance by 1 to 2\%.
  \item\textbf{DT}:  it had the lowest performance among all the included models. The rise in complexity of the decision tree's model will be much more costly than the other classifiers and even more than the increase in complexity of the system itself. In addition, the performance is not expected to match any of those costs. The rise in the dimensionality and the number of words is also expected to increase the overfitting risk.
\end{itemize}

		\subsection{Future work directions}
Several upgrades can be implemented -individually or a mix of them-:
\begin{itemize}
  \item Increase the diversity and size of the dataset to check the current limitations of the method in its existing configuration.
  \item The number of data inputs can be decreased to time-series arrays of the same number as sensors used, allowing the use of recurrent models (most notably: the Long Short Term Memory networks). In such a case classification of signs on-the-fly in a continuous way is expected to be possible
  \item The densely recorded data samples can be treated like image pixel values and fed into Convolutional Neural Network (CNN).
  \item Influence of adding additional EMGs, PPGs, or IMUs may allow efficient classification of signs that include twisting motions.
\end{itemize}

%%%%%%%%%%%%%%%%%%%%%%%%%%%%%%%%%%%%%%%%%%
\section{Conclusions}

This work presented a proof of concept of a new method of ASL interpretation. The proposed approach requires no predetermined environment or expensive equipment and is expected to be operational using a portable device. The method uses data acquisition in the form of arrays of instances and considers the sequence of those instances. High-quality performance in predicting new data was achieved as a result.\\

The highest overall performance, reaching up to 99\%, was achieved for RF and SVM classifiers. Moreover, both of them are expected to scale well in case a system or task complexity increases. The KNN models, while reaching relatively high performance (98\%), are not expected to work well in the long run. The DT and LR methods achieved subpar performance and, for this reason, are not found capable of solving this task in the future.

The obtained results can serve as a starting point for large-scale research in sign language interpretation, opening new possibilities for scalability and availability of such systems.

%%%%%%%%%%%%%%%%%%%%%%%%%%%%%%%%%%%%%%%%%%
\section*{Contribution}
Conceptualization, B.K. and Z.D.; methodology, B.K.; software, B.K.; validation, B.K.; formal analysis, B.K and Z.D..; investigation, B.K.; resources, B.K. and Z.D.; data curation, B.K.; writing---original draft preparation, B.K.; writing---review and editing Z.D.; visualization, B.K.; supervision, Z.D.; project administration, Z.D.; funding acquisition, Z.D. All authors have read and agreed to the published version of the manuscript.

\section*{Acknowledgement}

The work was funded by the research subsidy of the Department of Robotics and Mechatronics

Authors would like to express their gratitude to Abdullah Tarek and Aliaa Tahre for providing technical support and to Maria Leszko, Klaudia Kaczor and Aleksandra Soja for volunteering in the data aquisition testing phase

\singlespacing
\scriptsize
\bibliography{BK,ZD}
\label{sec:ref}
%\bibliography{LA_bib}
\bibliographystyle{ieeetr}

\end{document}